# Charge Carrier Mediation and Ferromagnetism induced in MnBi$_6$Te$_{10}$ Magnetic Topological Insulators by antimony doping

*Hangkai Xie*[1,2], *Fucong Fei*[1,2]\*, *Fenzhen Fang*[1]\*, *Bo Chen*[1,2], *Jingwen Guo*[1,2], *Yu Du*[1,2], *Wuyi Qi*[1,2], *Yufan Pei*[1,2], *Tianqi Wang*[1,2], *Muhammad Naveed*[1,2], *Shuai Zhang*[1,2], *Minhao Zhang*[1,2], *Xuefeng Wang*[2,3], *Fengqi Song*[1,2]

[1] *National Laboratory of Solid State Microstructures, Collaborative Innovation Center of Advanced Microstructures, and College of Physics, Nanjing University, 210093 Nanjing, China.*

[2] *Atomic Manufacture Institute (AMI), 211805 Nanjing, China.*

[3] *National Laboratory of Solid State Microstructures, Collaborative Innovation Center of Advanced Microstructures, and School of Electronic Science and Engineering, Nanjing University, 210093 Nanjing, China.*

\*Correspondence and requests for materials should be addressed to Fucong Fei (email: feifucong@nju.edu.cn) and Fenzhen Fang (email: fenzhen@nju.edu.cn).

**Abstract:**

A new kind of intrinsic magnetic topological insulators (MTI) MnBi$_2$Te$_4$ family have shed light on the observation of novel topological quantum effect such as quantum anomalous Hall effect (QAHE). However, the strong anti-ferromagnetic (AFM) coupling and high carrier concentration in the bulk hinder the practical applications. In closely related materials MnBi$_4$Te$_7$ and MnBi$_6$Te$_{10}$, the interlayer magnetic coupling is greatly suppressed by Bi$_2$Te$_3$ layer intercalation. However, AFM is still the ground state in these compounds. Here by magnetic and transport measurements, we demonstrate that Sb substitutional dopant plays a dual role in MnBi$_6$Te$_{10}$, which can not only adjust the charge carrier type and the concentration, but also induce the solid into a ferromagnetic (FM) ground state. AFM ground state region which is also close to the charge neutral point can be found in the phase diagram of Mn(Sb$_x$Bi$_{1-x}$)$_6$Te$_{10}$ when $x \sim$ 0.25. An intrinsic FM-MTI candidate is thus demonstrated, and it may take a step further for the realization of high-quality and high-temperature QAHE and the related topological quantum effects in the future.

**Introduction**

The novel topological quantum effects such as quantum anomalous Hall effect (QAHE), topological axion state, and the related topological electromagnetic effect can be achieved when combine the magnetic property to topological insulators (TIs) [1-4]. However, from the first realization of QAHE in Cr-doped $(Bi,Sb)_2Te_3$ in 2013 [5], the observation temperature is still extremely low, which is limited by extrinsic magnetic elemental dopants or the magnetic proximity heterostructures in magnetic topological insulators (MTIs) [6-9]. Recently, a new class of MTI-$MnBi_2Te_4$/$(Bi_2Te_3)_n$ family ($n$ = 0, 1, 2..) materials is discovered which is proposed to hold intrinsic magnetic property instead of the external magnetic impurities [10-16]. Because of the strong anti-ferromagnetic (AFM) coupling between $MnBi_2Te_4$ septuple layers (SLs), experimentally observed QAHE in $MnBi_2Te_4$ needs assistance of a large external magnetic field (~ 6 to 8 T) to form a ferromagnetic (FM) state in $MnBi_2Te_4$ [17-20], or with the extremely high bias voltage up to 200 V in a five SLs $MnBi_2Te_4$ device under zero field [21]. Fortunately, the interlayer AFM exchange interaction can be significantly reduced by intercalated $Bi_2Te_3$ quintuple layers (QLs) into $MnBi_2Te_4$ matrix [22-28], the $MnBi_4Te_7$ ($n$ = 1) can significantly reduce the AFM coupling[25, 29-31] and then the $MnBi_6Te_{10}$ ($n$ = 2) can almost hold fully FM state at zero field [22, 23, 28, 32], but the AFM state still possesses the lower energy than FM state as ground state so that a small field is still essential to overcome AFM coupling. Furthermore, in all of these intrinsic MTI candidates $MnBi_2Te_4$/$(Bi_2Te_3)_n$($n$ = 0, 1, 2..), the heavy n-type trivial bulk carriers will be induced because of the ineluctable antisite defects between the manganese and bismuth atoms, hindering the novel transport signatures contributed by the chiral edge state located in the exchange gap opening at the Dirac point of the surface states. The further applications of $MnBi_2Te_4$ and the related intercalated materials in the spintronics and dissipationless quantum devices are also obstructed.

In $MnBi_2Te_4$, the Sb substitution at the Bi sites can neutralize the excess bulk carrier efficiently [33-35], while the AFM coupling is still robust. On the other hand, Sb substitution in $MnBi_4Te_7$ can tune the excess bulk carrier and adjust the magnetic interaction between MnBi$_2$Te$_4$ SLs at the same time [36, 37], but only a subtle FM state can be achieved when Fermi level aligns with a Van Hove singularity in the bulk conduction band [38], no truly FM-MTI can be obtained. Fortunately, in the case of MnBi$_6$Te$_{10}$ discussed in this work, the AFM coupling can be completely overcome and restrain the bulk carrier concentration at the meantime by Sb substitution at the Bi sites, with no-magnetic elements doping, paving the way to demonstrate a real intrinsic FM-TI candidate for QAHE. A serials single crystals of Mn(Sb$_x$Bi$_{1-x}$)$_6$Te$_{10}$ from $x = 0$ to $x = 0.32$ are grown. Then the magnetic and electrical transport measurement are conducted to explore the influence of the Sb substitution in MnBi$_6$Te$_{10}$. From the magnetic measurement, we find the FM state takes the place of AFM state as ground state in Mn(Sb$_x$Bi$_{1-x}$)$_6$Te$_{10}$, even in a low level of Sb substitution when $x = 0.11$, implying the AFM coupling between SLs is been completely overcome, which is also consistent with previous report about the magnetic measurement in Sb element doped MnBi$_6$Te$_{10}$[39]. Furthermore, according to the electrical transport, similar to the case of MnBi$_2$Te$_4$ and MnBi$_4$Te$_7$, one can realize that the Sb element doping is also an effective approach to modulate the bulk carrier densities. Specifically, the heavy n-type bulk carriers in pure MnBi$_6$Te$_{10}$ can be significantly suppressed by Sb doping, and the charge neutral point (CNP) in n-p transition process appears near $x = 0.25$, which is also in the FM state region with fully net magnetization at zero field. Our work demonstrates the engineering of the both magnetic property and the charge carriers in MnBi$_6$Te$_{10}$ by Sb doping, and Mn(Sb$_x$Bi$_{1-x}$)$_6$Te$_{10}$ is a promising candidate of FM-TI and low bulk carrier concentrations under certain doping ratio, which provides a prospective avenue to accomplish the QAHE at the zero field as well as other exotic topological effects.

**Results**

Mn(Sb$_x$Bi$_{1-x}$)$_6$Te$_{10}$ is a layered rhombohedral material with the space group $R\bar{3}m$ [22, 28, 39, 40] stacked with one SL of Mn(Sb$_x$Bi$_{1-x}$)$_2$Te$_4$ and two QLs of (Sb$_x$Bi$_{1-x}$)$_2$Te$_3$ alternately through the weak van der Waals forces along the $c$ axis as seen in Fig. 1(a). The high quality crystals of Mn(Sb$_x$Bi$_{1-x}$)$_6$Te$_{10}$ are obtained by self-flux method [41].

The molar ratio of reactants is around MnTe: (Sb$_2$Te$_3$+Bi$_2$Te$_3$) = 1:8, then put it an alumina crucible and sealed by a quartz tube before put into a muffle furnace. Temperature rises up to 950℃ in 24 hours and maintains at this temperature for 6 hours, then slowly cools down to around 580℃, followed by centrifugation to separate the lustrous flakes from the excess flux, the detailed molar ratio of reactants and centrifugation temperature for every different Sb doped sample can be seen in the Table.S1 in the Supplementary. The inset of the Fig. 1(b) is the morphology image of the as-grown Mn(Sb$_x$Bi$_{1-x}$)$_6$Te$_{10}$ crystals. Fig. 1(c) displays the energy dispersive spectra (EDS) which shows the stoichiometric ratio of Mn: (Bi + Sb): Te is close to 1: 6: 10, the concrete composition ratio measured by EDS for every sample can be seen in Table.S1 in the Supplementary. The relative intensity of Bi at 3.8 keV and Sb at 3.6 keV evolve as expected after normalizing the counts with the Te peaks, elucidating the valid substitution of the two elements. The real ratio between Mn and Bi is deviated from the stoichiometric ratio a little bit, and this is believed to be the anti-site defects of Mn and Bi(Sb) [42-44]. Then we check the single crystal X ray diffraction (XRD) patterns for every Sb doped sample which show prominent peaks labeling (00$n$) Miller indexes in Fig. 1(b), and one can see all the XRD patterns in Fig. 1(b) only show a pure MnBi$_6$Te$_{10}$ phase pattern without intermixing other phase pattern from MnBi$_4$Te$_7$ or MnBi$_2$Te$_4$ or some other phases [22, 28, 39], which mean the Sb element has entered into the MnBi$_6$Te$_{10}$ structure introducing no other composition. The extracted value of the $c$ axis from XRD pattern is nearly at ~ 102 Å with just tiny change at different Sb doped ratio, as listed in Table.S1 in the Supplementary for every sample, indicating the similar structure constants and identical layered crystal structure for different Sb doped Mn(Sb$_x$Bi$_{1-x}$)$_6$Te$_{10}$ samples[39]. Both the XRD and EDS results demonstrate the favourable crystallinity and the precise molar ratio during the procedure of the substitution in Mn(Sb$_x$Bi$_{1-x}$)$_6$Te$_{10}$.

To determine the magnetic properties in the Mn(Sb$_x$Bi$_{1-x}$)$_6$Te$_{10}$, magnetic measurements by a vibrating sample magnetometer equipped on a physical property measurement system (Quantum Design PPMS-14T) are carried out. First we measure the field cool (FC) and zero-field cool (ZFC) processes from 300 K to 2 K for each different Sb doped sample, as is shown in Fig. 2(a). The magnetic field is along $c$ axis of the sample cleave surfaces with the magnitude at 500 Oe. In the pure $MnBi_6Te_{10}$, the FC-ZFC curves show a clear typical antiferromagnetic peaks at 10.1 K and the overlapped ones diverge at lower temperature (~ 5 K), which is corresponding with previous reports [22, 32]. The magnetic field dependence of the magnetization curves in Fig. 2(b) exhibit the zigzag shape and the hysteresis loop divides into two jump points deviating from the zero field at around ±0.07 T obviously from 2 K to 8 K, which also indicates the existence AFM coupling. Although the saturation field to approach the FM region in the $MnBi_6Te_{10}$ is just around 0.2 T, which is much smaller than the one in $MnBi_2Te_4$ (6~8 T), the magnetic ground state in $MnBi_6Te_{10}$ is still AFM state and external magnetic field is still essential to form a FM state.

The $Bi_2Te_3$ QLs intercalation enlarges the distance between the neighboring $MnBi_2Te_4$ SLs, and the energy difference between the AFM and FM is getting much smaller from $MnBi_2Te_4$ to $MnBi_4Te_7$ then to $MnBi_6Te_{10}$ [32]. Weak interlayer AFM coupling paves the way for the magnetic property modulation by various approaches. In addition, when substituting Bi with the no-magnetic Sb element in $MnBi_6Te_{10}$, the AFM ground state can be completely eliminated without introducing the undesired outer magnetic impurities, and then replaced by FM ground state. As seen in Fig. 2(a), the FC-ZFC curves of the all Sb doped samples show the λ shaped typical ferromagnetic properties. Then we can get the transition temperature by differentiating the magnetization versus temperature curves, it can show clear kinks around 12 K in all Sb doped samples, indicating the Curie temperature ($T_c$) [25], which are also slightly higher than the Néel temperature ($T_N$) in pure $MnBi_6Te_{10}$, as is shown in the Fig. 2(a). Furthermore, the magnitudes of the divergence between FC-ZFC curves in Sb doped samples at lower temperature (~ 5 K) rise are much larger than the divergence value in pure sample with AFM transition point at 10.1 K, proving the observably enhancement of the FM ingredient compared with pure $MnBi_6Te_{10}$. It implies the Sb substitution at Bi site in $MnBi_6Te_{10}$ is an effective way to modulate the magnetic interaction between SLs from AFM to FM state.

To further explore the evolution of magnetic property of the various Sb doping ratios, magnetization versus magnetic field (*M-H*) measurement is conducted shown in Fig. 2(b)-2(g). In pure MnBi$_6$Te$_{10}$, one can see the feature of zigzag shaped hysteresis loops at different temperature and evolves to the inclined rectangle shaped loops at around ±0.07 T in Fig. 2(b). Of course one can notice at 2 K, the zigzag shaped hysteresis loops merged to be one hysteresis loop to a large extent, confirming the interlayer AFM exchange coupling between MnBi$_2$Te$_4$ SLs is significantly weakened by the intercalating Bi$_2$Te$_3$ QLs, but there is still the main domination of the AFM feature in MnBi$_6$Te$_{10}$, especially at higher temperature above 2 K. On the other hand, even by a low level Sb doping ($x = 0.11$), the zigzag shape and the hysteresis loop separation in *M-H* curves are completely gone, as is shown in the Fig. 2(c). The hysteresis loop demonstrate one pure large inclined rectangle shaped loop with fully polarized FM state stabilized at zero fields at 2 K, which is beneficial to realize the QAHE under zero field. In addition, when increasing the temperature, the loop does not split into two separated parts with zigzag shaped curve but just simply shrinks and is always pinned at the zero field. Along with Sb doping ratio increasing continuously, the FM state at Mn(Sb$_x$Bi$_{1-x}$)$_6$Te$_{10}$ is still holding, and all of them are capable of holding fully polarized FM state at zero fields at 2 K from $x = 0.11$ to $x = 0.25$, as shown in Fig. 2(c)-2(f). When $x$ rises up to around 0.32, the polarized FM state cannot be fully maintained at zero fields even at 2 K, as is shown in Fig. 2(g). Although the magnetic transition temperature is rising by Sb doping, the coercive field is dropping down to around 300 Oe or even less for Sb doped samples. It may be explained by the effect of magnetic anisotropy caused by the introduction of Sb element impurities or some other different type of magnetic interaction appears which can compete with single FM interaction between SLs [39]. Anyway, the results can suggest the good FM state region can be hold at a Sb doping level range from $x = 0.11$ to $x = 0.25$ in Mn(Sb$_x$Bi$_{1-x}$)$_6$Te$_{10}$. This kind of hysteresis behavior demonstrates a typical FM ordering after Sb doping in MnBi$_6$Te$_{10}$, consistent with the FC-ZFC behavior discussed above. It can be believed that after Sb element doping, the Mn-Sb site mixing where Mn occupying the Sb site probably mediates a ferromagnetic coupling between Mn layers in MnBi$_2$Te$_4$ family in some recent researches, which will lead to the changing of energy between AFM and

FM states [35, 45-48].

Accompanying by the engineering of the magnetic property, Sb element doping can also modify the level of carrier density in $MnBi_6Te_{10}$. Due to the internal defects of Mn and Bi antisite, the as grown $MnBi_2Te_4/(Bi_2Te_3)_n$ ($n = 0, 1, 2..$) family are all serious n-type doped regardless of the growth methods [12, 43], which are also consistent with the previous ARPES measurement reports[15, 16, 23, 25, 28, 49, 50]. Heavy carriers in bulk will be a big obstacle for the capacity of regulating gate voltage at ultra-thin films in device running process. Fortunately, no-magnetic element doping is an efficient way to adjust the Fermi level of the samples without destructing the topological property and the carrier mobility according to several previous reports in these related materials [33-35, 51, 52]. We perform the electrical transport measurement in $MnBi_6Te_{10}$ after Sb doping including magneto and Hall resistivity measurements. As shown in Fig. 3(a), the resistivity versus temperature curves display characteristic kinks marked by arrows at around 12 K in $Mn(Sb_xBi_{1-x})_6Te_{10}$ for all different Sb doping concentration, and the kink of $x = 0$ sample is around 10 K, which are consistent with the magnetic transition temperature obtained from the magnetic measurement in Fig. 2(a). It is also noticeable that the resistivity firstly rises when increasing the Sb doping ratio, and reaches the maximum when $x = 0.25$. When further increasing the value of $x$ above 0.25, the resistivity drops. Distinct to the metallic behavior in other samples, the sample with $x = 0.25$ also seems to show a semiconductor behavior with a slight incensement of resistivity when temperature drops to around 200 K. Thus it suggests that the carrier concentration is suppressed and exhibits properties similar to semiconductors, which also indict that the Fermi level is close to the n-p transition point in $x = 0.25$ Sb doped sample. Then one can pay attention to the longitudinal and Hall resistivity as displayed in Fig. 3(b)-3(e), anomalous Hall effect (AHE) can be clearly seen in Sb doped $MnBi_6Te_{10}$ samples. For pure $MnBi_6Te_{10}$, one can see the AHE hysteresis curves display zigzag shaped loops and show two part of the inclined rectangle shaped loops at around ±0.07 T at 2 K in Fig. 3(b), as well as the butterfly-shaped hysteresis behavior in longitudinal resistivity as shown in Fig. 3(c). When temperature rises up to 5 K, the inclined rectangle shaped loops starts to separate into two isolated small loops, again confirming the AFM ground state in pure MnBi$_6$Te$_{10}$. Then let us shift the sight to the AHE signal in Sb doped samples, consistent with the magnetic measurements, the AHE signal can show typical FM-type shaped loops with fully polarized FM state stabilized at zero fields at 2 K. The loop simply shrinks and still performs as a form of FM state as the temperature goes up till $T_c$ from $x = 0.11$ to $x = 0.20$, as shown in Fig. 3(b). At the meantime, the longitudinal resistivity shows the clear butterfly curves at 2 K in Fig. 3(c), which further confirms the net magnetization emerging near zero fields. It is clear that the AHE signal hysteresis always maintains when $x = 0.11$ and $x = 0.2$. When Sb doping concentration is up to 0.25, the hysteresis loop cannot be clearly seen and the Hall signal show round S-shaped curves, which implies the Fermi level in Sb doped sample with $x = 0.25$ is close to the Dirac point, so the Hall measurement demonstrates a two-band feature, as is shown in Fig. 3(d). When $x$ rising up to 0.32, the FM-type AHE emerges again, with the hysteresis loops shrink compared with previous lower Sb doped samples and the fully AHE signal at zero field cannot be hold at 2 K, same as the magnetic measurement in Fig. 2(g). It is also worth noticing that the Hall signal completely changes the sign from negative to positive, which is similar to the ones in MnBi$_2$Te$_4$ series reported in the previous reports and can be attributed to the competition of intrinsic Berry curvature and extrinsic skew scattering [53, 54], as is shown in Fig. 3(b). This also explains the disappearing AHE signal hysteresis in Mn(Sb$_{0.25}$Bi$_{0.75}$)$_6$Te$_{10}$ sample, the two opposite AHE signals from two distinct origins coincidently cancel out each other. Then one can focus on the carrier densities by extracting the Hall coefficient according to the slope of the Hall measurements up to 9 T at 2 K for every Sb doped sample, as seen in Fig. S1 in the Supplementary. In pure MnBi$_6$Te$_{10}$, there is n-type carrier with a high concentration at around $2.68 \times 10^{20}$ cm$^{-3}$ at 2 K, similar to MnBi$_4$Te$_7$ and MnBi$_2$Te$_4$ with high n-type carrier concentration which also can show the metallic properties [22, 25, 33, 55-57]. After Sb doping, the carrier concentration is about $2.58 \times 10^{20}$ cm$^{-3}$ when $x = 0.11$ for Sb doping at 2 K, almost same with the pure MnBi$_6$Te$_{10}$. Then $x$ goes up to 0.17, the concentration is little lower at around $2.07 \times 10^{20}$ cm$^{-3}$, it is still a high bulk carrier density. When $x = 0.2$ for Sb doping, one can notice that the carrier concentration has dropped to around $8.07 \times 10^{19}$ cm$^{-3}$, much smaller than the pure sample, and it is still n-type carriers in bulk. As for the one with $x$ = 0.32, the carrier concentration is about $5.3\times10^{19}$ cm$^{-3}$. More importantly, the symbol of the Hall coefficient changes from negative to positive and the bulk carrier is p-type now, which means that the n-p transition occurs in this region, corresponding to the behavior in the Hall signal in $x$ = 0.25 Mn(Sb$_x$Bi$_{1-x}$)$_6$Te$_{10}$ sample, as shown in Fig. 3(d). Here we use the classical two-band magneto transport model [58] to fit Hall data at 2 K in $x$ = 0.25 Sb doped sample to extract two kind of carrier concentration, as seen in Fig. S2 in the Supplementary. The Hall resistivity can be described as

$$\rho_{yx} = \frac{(p\mu_h^2 - n\mu_e^2)B + \mu_h^2\mu_e^2(p-n)B^3}{e[(n\mu_e + p\mu_h)^2 + (p-n)^2\mu_e^2\mu_h^2B^2]} \quad (1)$$

, where $n$ ($p$) and $\mu_e$ ($\mu_h$) are the carrier density and mobility for electrons (holes), respectively, and $B$ is the magnetic field. Then we get $6.1\times10^{19}$ cm$^{-3}$ and $2.0\times10^{19}$ cm$^{-3}$ for n-type and p-type carrier density respectively, and both of them are much lower than the carrier density in pure MnBi$_6$Te$_{10}$ sample and other Sb doped samples, proving again the Fermi level in $x$ = 0.25 sample is around Dirac point and Sb element substitution at Bi site in MnBi$_6$Te$_{10}$ is an effective way to modulate the chemical potential. Furthermore, let us take more attention to the magneto transport behavior of the $x$ = 0.25 Sb doped sample, the longitudinal resistivity also shows the clear butterfly curves displayed in Fig. 3(e), same as the other Sb doped MnBi$_6$Te$_{10}$, indicting the spontaneous magnetization with FM ground state is also maintained in Mn(Sb$_{0.25}$Bi$_{0.75}$)$_6$Te$_{10}$ which is also corresponding to magnetic property in Mn(Sb$_{0.25}$Bi$_{0.75}$)$_6$Te$_{10}$ we measured above in Fig. 2(f). We also measured the transport signals for some other Sb doped MnBi$_6$Te$_{10}$ samples around $x$ = 0.25 for further step verification, the semiconductor-like behaviors are repeatable, and the carrier densities are also in a very low level compared to other different Sb ratio samples, as seen in Fig. S3-S5 in the Supplementary. Contrary to stubborn AFM in the Sb doped MnBi$_2$Te$_4$ [33] and complex competition between AFM and FM states in Sb doped MnBi$_4$Te$_7$ [38], it is apparent that the FM ground state can be approached accompanied with the successful modulation of carrier densities in Sb-doped MnBi$_6$Te$_{10}$ samples.

  To clearly elucidate the properties of the magnetism and the electrical transport during the procedure of Sb doping in MnBi$_6$Te$_{10}$, we summary the magnetic transition temperature and the charge carrier concentrations of the Mn(Sb$_x$Bi$_{1-x}$)$_6$Te$_{10}$ samples with various Sb doping ratio in the diagram displayed in Fig. 4. In pure MnBi$_6$Te$_{10}$, the n-type carrier concentration is high and totally in an AFM state region with a Néel temperature at 10.1 K. As the Sb doping ratio increases, Mn(Sb$_x$Bi$_{1-x}$)$_6$Te$_{10}$ quickly takes step into the FM ground state region and there is a slight rise to around 12 K for magnetic transition point. At the meantime, the bulk carrier density keeps falling continuously. More importantly, the n-type carrier concentration is significant reduced at the Sb doping ratio of 20%, and the bulk carrier change its sign being p-type at 32% ratio, implying the charge neutral point (CNP) is in this Sb ratio region. It can be explained by the reason that the Sb vacancies or Sb-on-Te antisites (Sb$_{Te}$) are incrementally introduced into Mn(Sb$_x$Bi$_{1-x}$)$_6$Te$_{10}$[59-61], then the n-type carrier is suppressed by Sb element substitution at Bi sites till $x \sim 0.25$. After Sb doping ratio is up to around 0.32, the Sb vacancies or Sb$_{Te}$ antisites take a remarkable effect for the p-type background. Furthermore, from the phase diagram one can clearly see that the CNP takes place deeply into the FM state region colored by the cyan background, ensuring the strong spontaneous magnetization with parallel magnetic moment arrangement in bulk carrier suppressed Mn(Sb$_x$Bi$_{1-x}$)$_6$Te$_{10}$ with $x$ around 0.25. It is also confirmed by our measurement combined with the magnetization and electrical measurement at $x = 0.25$ Sb doped sample, which combine both n-type and p-type carriers near the Dirac point, and two type carrier densities are both restrained at a relatively low level. Then we can demonstrate the real intrinsic ferromagnetic topological insulator in Mn(Sb$_x$Bi$_{1-x}$)$_6$Te$_{10}$, especially when $x = 0.25$, with the Fermi level is close to the Dirac point, which provide a promising platform to explore and realize the QAHE and other topological quantum effect.

**Conclusion**

In summary, the magnetic property and the carrier density have been modulated simultaneously in the Mn(Sb$_x$Bi$_{1-x}$)$_6$Te$_{10}$, The AFM ground state have evolved into FM state after Sb doping, and always keep in the FM state region at different doping ratio, the AFM coupling between SLs has been completely suppressed. Besides, the dominated charge carriers can be modified from n-type to p-type effectively, and charge neutral point with greatly suppressed bulk carrier concentrations can be achieved in the sample with Sb ratio around $x = 0.25$. Meanwhile, these samples close to the charge neutral point also hold the FM ground state at zero external field, demonstrating a promised candidate of MTI with FM ground state and lower bulk carrier concentrations. The intriguing magnetic and charge carriers transitions induced by Sb dopants also provide an excellent avenue to study the interaction coupling between the magnetism and carriers in MTIs. And the most important thing is that $Mn(Sb_xBi_{1-x})_6Te_{10}$ displays improved performance on both magnetic property and electrical transport than pure $MnBi_2Te_4/(Bi_2Te_3)_n$ series, shedding light to the exploration and realization long-expected QAHE and the related quantum topological effect.

**Acknowledgments:**

The authors gratefully acknowledge the financial support of the National Key R&D Program of China (2017YFA0303203); the National Natural Science Foundation of China (12025404, 91622115, 11522432, 11574217, U1732273, U1732159, 61822403, 11874203, 11904165, and 11904166); the Natural Science Foundation of Jiangsu Province (BK20190286); the Fundamental Research Funds for the Central Universities (020414380150, 020414380151, 020414380152, 020414380192, 020414380194); the Users with Excellence Project of Hefei Science Center, CAS (Grant No. 2019HSC-UE007); and the opening Project of Wuhan National High Magnetic Field center.

**Figure captions**

**Figure. 1 Crystal growth and characterization of the Mn(Sb$_x$Bi$_{1-x}$)$_6$Te$_{10}$.** **(a)** The crystal structures of the Mn(Sb$_x$Bi$_{1-x}$)$_2$Te$_4$/(Sb$_x$Bi$_{1-x}$)$_2$Te$_3$ superlattice. **(b)** Single crystal X-ray diffraction of the five samples. The curves are offset for clarity. Inset: Optical images of the as-grown sample, the scale bar is 2 mm **(c)** Energy dispersive spectra after normalized with the intensity of the Te element.

**Figure. 2 Magnetic measurement of the Mn(Sb$_x$Bi$_{1-x}$)$_6$Te$_{10}$.** **(a)** FC-ZFC curves of the different Sb doped ratio samples when field is 500 Oe along $c$ aixs and the Néel temperature ($T_N$) or Curie temperature ($T_c$) for each sample. The curves are offset for clarity. **(b)-(g)** The magnetic field dependence of the magnetization from 2 K to 15 K when field is along $c$ aixs.

**Figure. 3 Electrical transport of the Mn(Sb$_x$Bi$_{1-x}$)$_6$Te$_{10}$.** **(a)** Resistivity versus temperature from 2 K to 300 K with the obvious kinks mark by arrows. **(b)** Hall and anomalous Hall (after subtracting the background) signal at the selected ratios: $x$ = 0.00, 0.11, 0.20 and 0.32. **(c)** Magneto resistivity signal at the selected ratios: $x$ = 0.00, 0.11, 0.20 and 0.32. **(d)** Hall signal at $x$ = 0.25 Sb doping sample. **(e)** Magneto resistivity signal at $x$ = 0.25 Sb doping sample.

**Figure. 4 Evolution diagrams of the magnetic property and the Carrier density.** The red star points represent the carrier density versus Sb doping concentration in Mn(Sb$_x$Bi$_{1-x}$)$_6$Te$_{10}$ at 2 K. The blue diamond points represent the magnetic phase transition temperature. Three different colors represent the magnetic states, Antiferromagnetic state (AFM), ferromagnetic state (FM) and Paramagnetic state (PM). The charge neutral point (CNP) is marked.

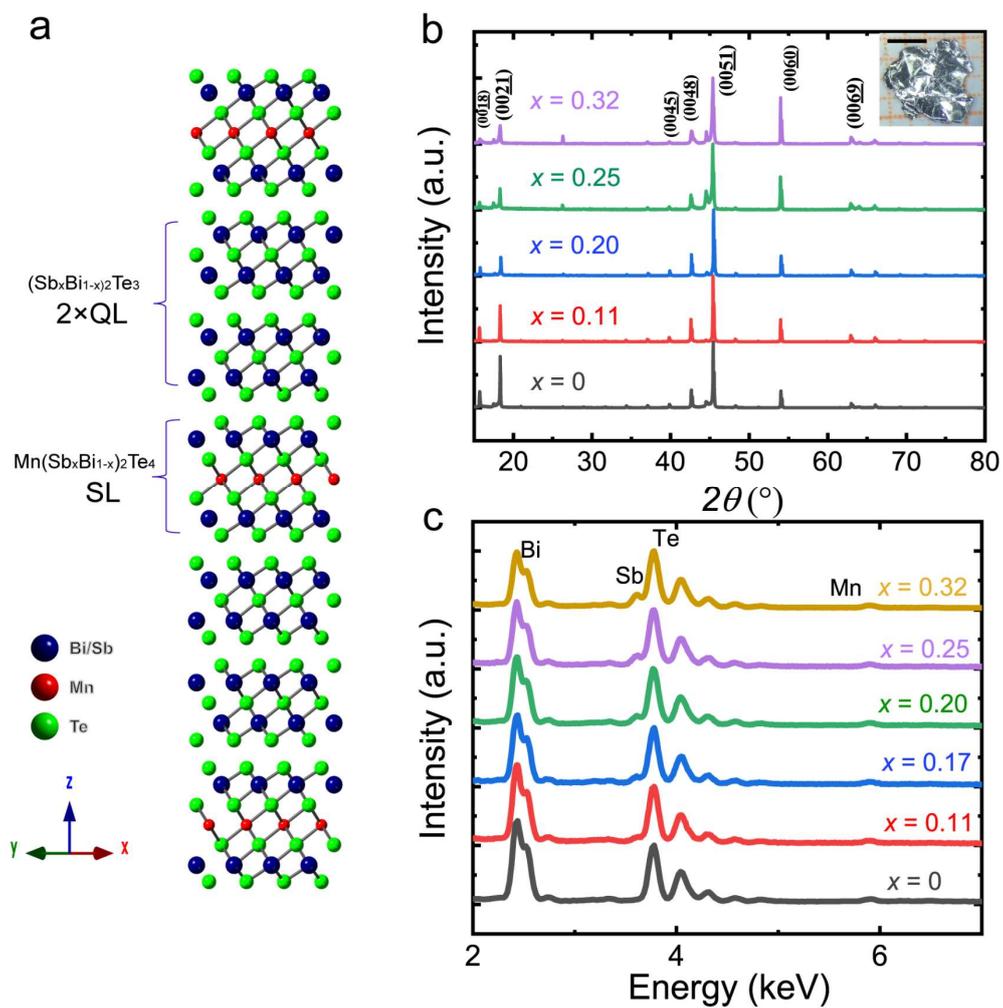

**Figure. 1 Crystal growth and characterization of the Mn(Sb$_x$Bi$_{1-x}$)$_6$Te$_{10}$.**

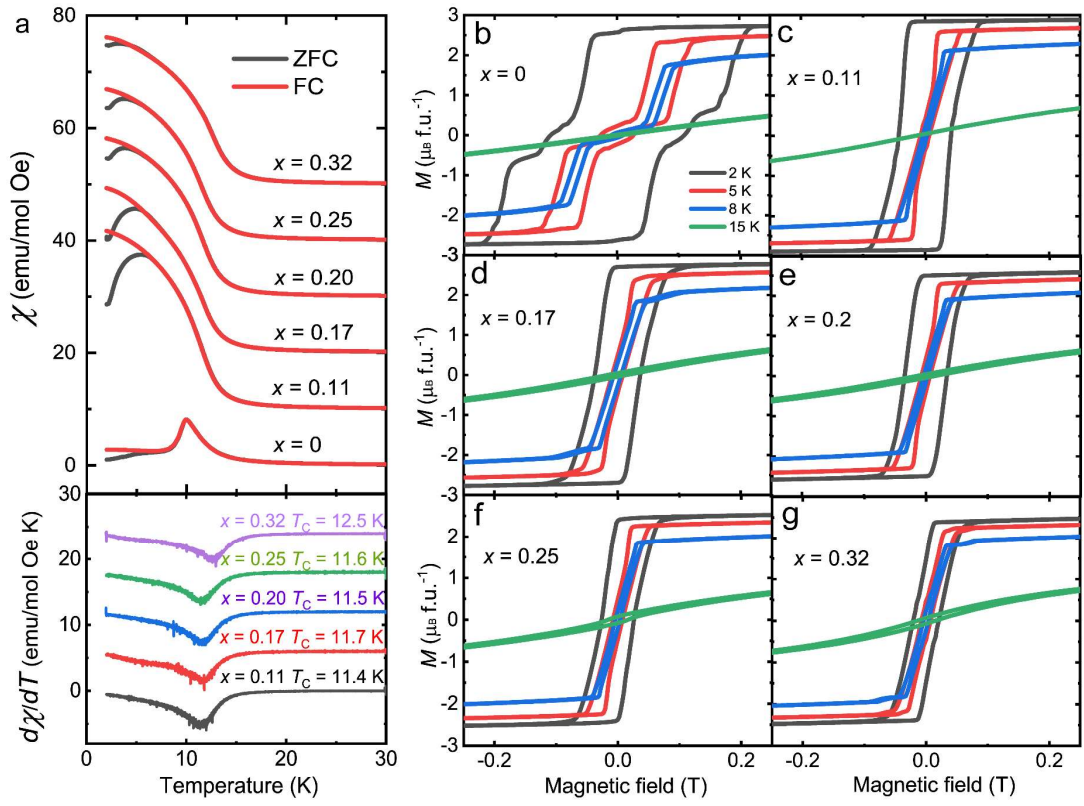

**Figure. 2 Magnetic measurement of the Mn(Sb$_x$Bi$_{1-x}$)$_6$Te$_{10}$.**

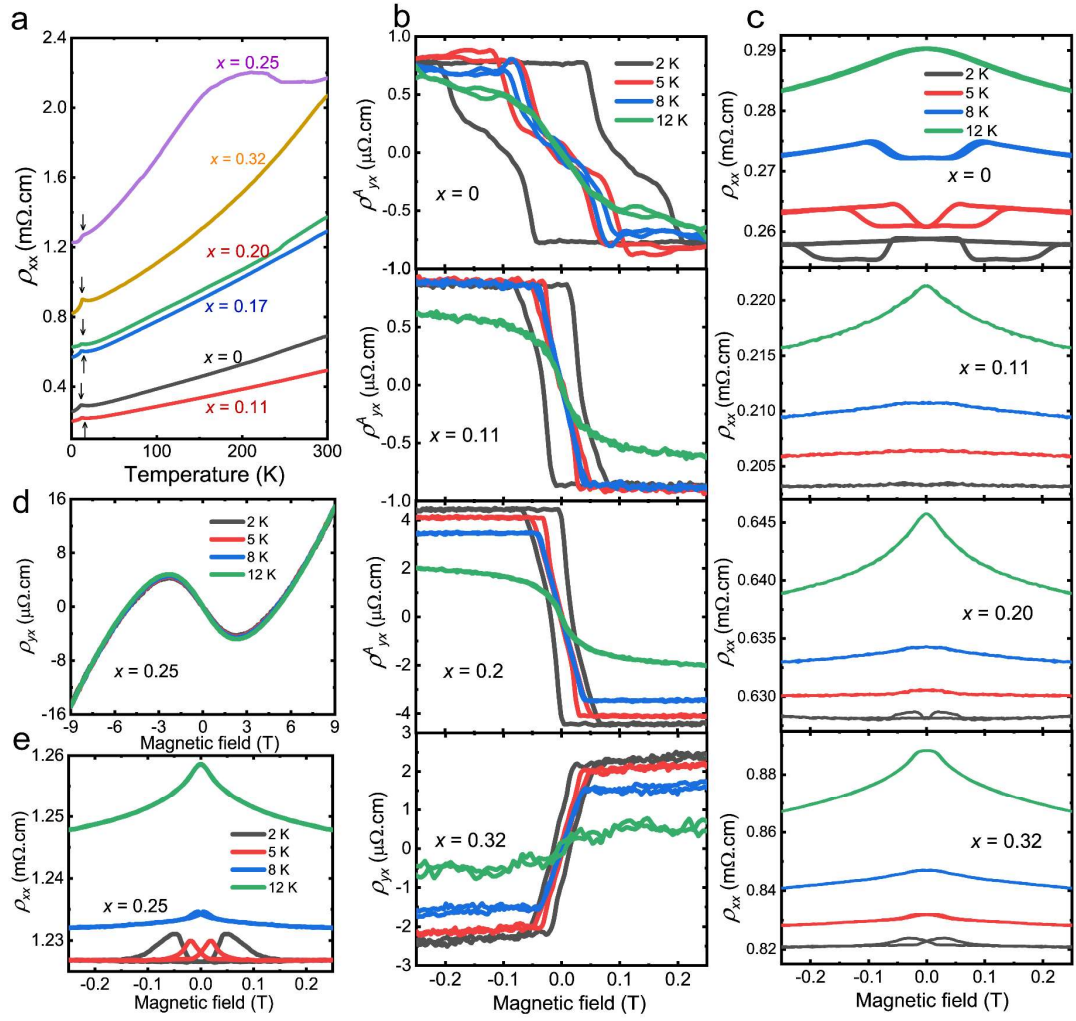

**Figure. 3 Electrical transport of the Mn(Sb$_x$Bi$_{1-x}$)$_6$Te$_{10}$.**

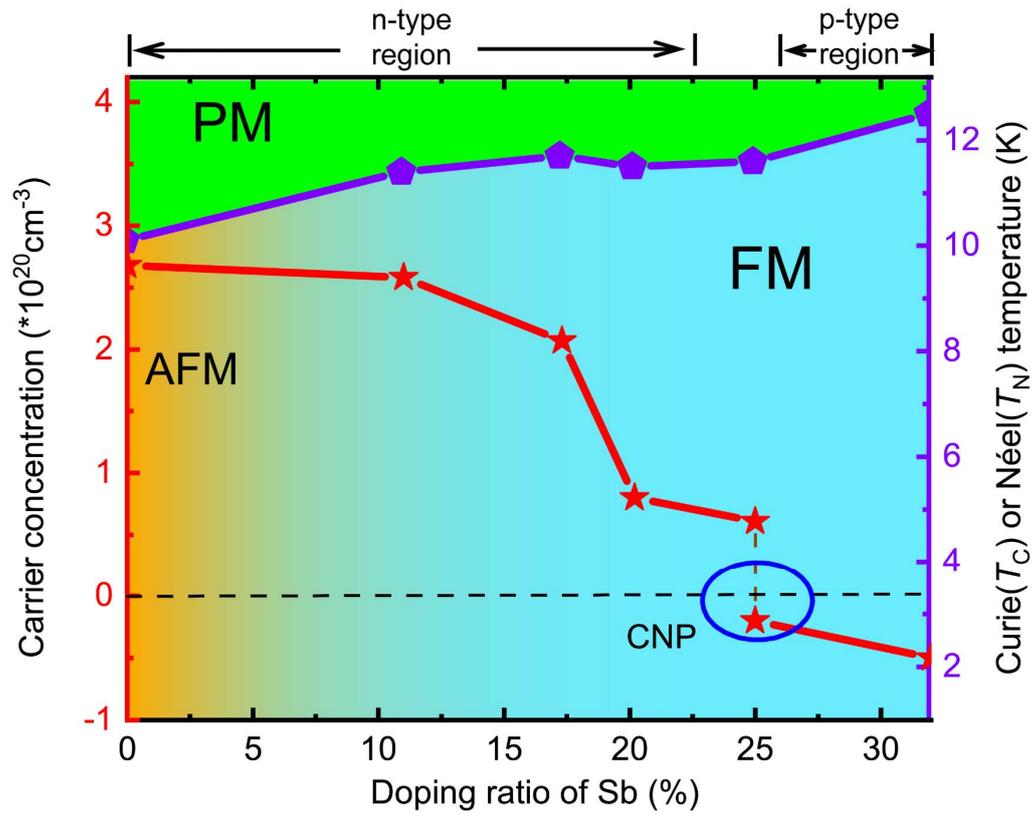

**Figure. 4 Evolution diagrams of the magnetic property and the Carrier density.**